\documentclass{article}
\pdfoutput=1
\usepackage{spconf,amsmath,graphicx}

\usepackage{color}
\usepackage{enumitem}
\usepackage{multirow}
\usepackage{etoolbox,siunitx}
\robustify\bfseries
\usepackage{booktabs}
\usepackage{balance}
\usepackage{amssymb}
\usepackage{bm}
\usepackage{arydshln}
\usepackage{caption}
\usepackage[labelformat=simple]{subcaption}

\usepackage{cite}
\usepackage[bookmarks=true,hypertexnames=true,pagebackref=false]{hyperref}
\hypersetup{
  colorlinks=true, % Colors links instead of ugly boxes
  urlcolor=black,   % Color for external hyperlinks
  linkcolor=blue,  % Color of internal links
  citecolor=red    % Color of citations
}

\usepackage{amssymb}% http://ctan.org/pkg/amssymb
\usepackage{pifont}% http://ctan.org/pkg/pifont
\newcommand{\cmark}{\ding{51}}%
\newcommand{\xmark}{\ding{55}}%

\def\T{{\mathsf T}}

\def\RR{{\mathbb R}}

\usepackage{caption}
\let\OLDthebibliography\thebibliography
\renewcommand\thebibliography[1]{
  \OLDthebibliography{#1}
  \setlength{\parskip}{0.5pt}
  \setlength{\itemsep}{1pt plus 0.3ex}
}
\newcommand{\finalsisdrtwo}{24.0}
\newcommand{\finalsisdrthree}{23.7}

\usepackage{flushend}

\title{Boosting Unknown-number Speaker Separation with\\Transformer Decoder-based Attractor}
\name{Younglo Lee$^1$, Shukjae Choi$^1$, Byeong-Yeol Kim$^1$, Zhong-Qiu Wang$^2$, Shinji Watanabe$^2$}
\address{
$^1$42dot Inc., Seoul, Korea \\
$^2$Language Technologies Institute, Carnegie Mellon University, Pittsburgh, USA \\
{\small\texttt{younglo.lee@42dot.ai}}
\vspace{-0.3cm}
}
\begin{document}
\ninept
%

%\titlespacing*{\section}{0pt}{2.ex plus 0ex minus 0ex}{0.5ex plus 0ex minus 0ex}
%\titlespacing*{\subsection}{0pt}{2.ex plus 0ex minus 0ex}{0.5ex plus 0ex minus 0ex}
%\titlespacing{\section}{0pt}{\parskip}{-\parskip}
%\titlespacing{\subsection}{0pt}{\parskip}{-\parskip}
%\titlespacing{\subsubsection}{0pt}{\parskip}{-\parskip}

\maketitle
\setlength{\abovedisplayskip}{2pt}
\setlength{\belowdisplayskip}{2pt}
\begin{abstract}
We propose a novel speech separation model designed to separate mixtures with an unknown number of speakers.
The proposed model stacks 1) a dual-path processing block that can model spectro-temporal patterns, 2) a transformer decoder-based attractor (TDA) calculation module that can deal with an unknown number of speakers, and 3) triple-path processing blocks that can model inter-speaker relations.
% The proposed TDA module extends an existing encoder-decoder attractor calculation module by replacing its LSTM layers with transformer decoder layers.
%Given a fixed, small set of learned speaker queries, TDA infers the relations of these queries and the mixture context produced by the dual-path blocks and directly generates individual attractor vectors for each speaker.
Given a fixed, small set of learned speaker queries and the mixture embedding produced by the dual-path blocks, TDA infers the relations of these queries and generates an attractor vector for each speaker.
The estimated attractors are then combined with the mixture embedding by feature-wise linear modulation conditioning, creating a speaker dimension.
The mixture embedding, conditioned with speaker information produced by TDA, is fed to the final triple-path blocks, which augment the dual-path blocks with an additional pathway dedicated to inter-speaker processing.
The proposed approach outperforms the previous best reported in the literature, achieving \finalsisdrtwo~and \finalsisdrthree~dB $\Delta\text{SI-SDR}$ on WSJ0-2 and 3mix respectively, with a single model trained to separate 2- and 3-speaker mixtures.
The proposed model also exhibits strong performance and generalizability at counting sources and separating mixtures with up to 5 speakers.
\end{abstract}
\section{Introduction}
\vspace{-.2cm}
Speech separation (SS) is the task of separating concurrent speech sources in a mixture signal recorded when multiple speakers talk simultaneously into individual speech signals.
Although humans exhibit a remarkable capability of perceiving any target sound of interest from a mixture of multiple sound sources, researchers have found it challenging to enable machines to have the same hearing capability, especially in monaural conditions~\cite{Qian2018past}.
%when only a single channel of the mixed signal is present~\cite{Qian2018past}.
Thanks to the strong modeling capabilities of deep neural networks (DNN), significant advances have been made in SS.
Since deep clustering \cite{Hershey2016, Isik2016, Wang2018AlternativeObejectives}, deep attractor network \cite{Chen2017deep, Luo2018speaker}, and permutation invariant training (PIT) \cite{Yu2017permutation, Kolbak2017} successfully solved the label-permutation problem in talker-independent speaker separation, many subsequent studies \cite{Luo2018TasNet, Luo2019conv, Luo2020, Chen2020DPTnet, Zeghidour2020, Subakan2021, Rixen2022QDPN, Wang2023TFGridNet, Wang2023TFGridNet2, Zhao2023Moss} have developed more efficient DNN architectures to improve the separation performance.
Through years of effort, the current state-of-the-art performance on the WSJ0-2mix dataset \cite{Hershey2016, Isik2016}, which is a popular benchmark designed for two-speaker separation, has reached an impressive scale-invariant signal-to-distortion ratio (SI-SDR) improvement of $23.5$ dB over the mixture~\cite{Wang2023TFGridNet, Wang2023TFGridNet2}.

%However, in most instances, it is commonly assumed that the number of speakers within the mixture is predetermined and remains unchanged, 
In many early studies, it is commonly assumed that the number of speakers within each mixture is known and fixed, which drastically limits the application range of their developed systems. In order to overcome this limitation, \cite{Kinoshita2018, ijcai2018p605, Takahashi2019, NEURIPS2020_27059a11, Shi2020} have proposed recursive multi-pass source separation schemes. Meanwhile, \cite{Nachmani2020, Zhu2021, Chazan2021} have proposed alternative approaches by training individual components for all conceivable numbers of speakers and at run time selecting one of the trained components, corresponding to the right number of speakers, to produce target estimates. However, their works require extra speaker identity information~\cite{ijcai2018p605, Shi2020}, recursive separation steps~\cite{Kinoshita2018, Takahashi2019, NEURIPS2020_27059a11}, or multiple models~\cite{Nachmani2020, Zhu2021, Chazan2021}.
Recently, encoder-decoder attractor (EDA) calculation-based methods have emerged, which are originally designed to handle an unknown number of speakers in speaker diarization \cite{Horiguchi2020end}, and demonstrated strong performance in SS~\cite{Chetupalli2022, Maiti2023EENDSS, Chetupalli2023}. 

Inspired by the above successful achievements, we propose SepTDA, a speech separation model that combines the strengths of the triple-path approach~\cite{Chetupalli2022, Chetupalli2023} and an LSTM-augmented self-attention block (denoted as LSTM-attention), which %enables to
can more efficiently capture local and global contextual information. In addition, a TDA calculation module is proposed with modifications to the original EDA~\cite{Horiguchi2020end} to effectively handle the undetermined count of speakers. We show that SepTDA outperforms the previous best reported in the literature by achieving state-of-the-art results on all the WSJ0-mix benchmark datasets \cite{Hershey2016, Isik2016, Nachmani2020}, which consist of mixtures with two to five speakers.
We share the separation samples\footnote{\scriptsize{\url{https://42speech.github.io/septda}}}.

\section{Proposed Method}\label{proposed}
\begin{figure}
\centering
\includegraphics[width=0.45\textwidth]{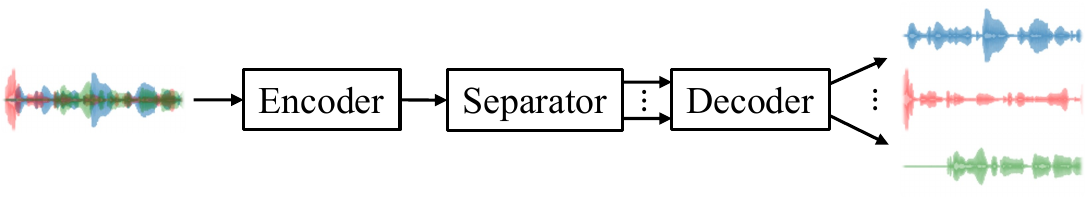}
%\caption{High-level illustration of our SS model pipeline.}
% \vspace{-1.5em}
\vspace{-0.2cm}
\caption{High-level system illustration of SepTDA.}
\label{fig:0}
\vspace{-0.4cm}
\end{figure}

\vspace{-.2cm}
The proposed SepTDA, illustrated in Fig.~\ref{fig:0}, is based on a time-domain encoder-decoder separation framework \cite{Luo2019conv}.
%for solving the cocktail party problem.
%We consider a mixture ${x(t)}$ of ${C}$ sources, ${y_c(t)}$ where ${c\in{\{1,\dots,C\}}}$.
Given a time-domain $T$-sample mixture ${\textbf{x}=\sum_{c=1}^{C}{\textbf{y}_c}}\in \RR^{T}$ with $C$ speakers, a separation model is trained to estimate each speaker source ${\textbf{y}_c}$.
SepTDA consists of an encoder, a separator, and a decoder.
This section describes each of them.
%Each stage is described in the following sub-sections.

% \vspace{-.3cm}
\vspace{-.1cm}
\subsection{Encoder}
\label{sec:encoder}
\vspace{-.1cm}

The encoder is a one-dimensional (1D) convolutional layer with a kernel size of $L$ samples and a stride size of $L/2$ samples.
It maps the mixture signal 
%$\textbf{x}=[x(1);\cdots;x(T)]\in{\mathbb{R}^{T}}$
$\textbf{x}$ to a $D_{e}$-dimensional latent representation $\textbf{E}\in{\mathbb{R}^{T' \times D_{e}}}$, where $T'=\left \lceil 2\times T/L \right \rceil$ with proper zero-padding:
\begin{align}
\label{eq:1}
\textbf{E}=\text{GELU}(\text{Conv1D}(\textbf{x})),
\end{align}
where $\text{GELU}(\cdot)$ denotes the Gaussian error linear unit activation function~\cite{Hendrycks2016Gaussian}.

\subsection{Separator}
\label{sec:separator}
\begin{figure}
\centering
\begin{subfigure}{0.48\textwidth}
\includegraphics[width=\textwidth]{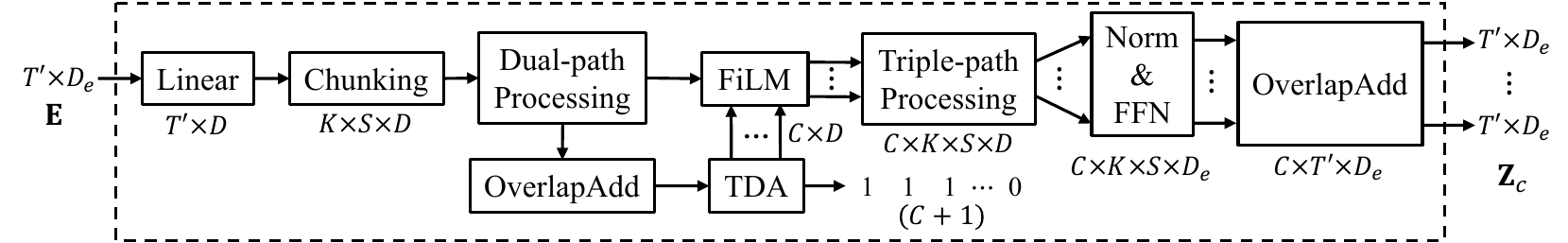}
\caption{Separator architecture.}\label{fig:2a}
\vspace{1em}
\end{subfigure}
\begin{subfigure}{0.4\textwidth}
\includegraphics[width=\textwidth]{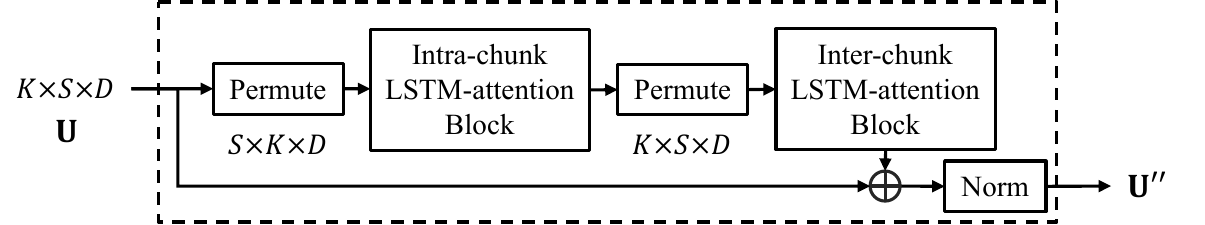}
\caption{Dual-path processing block.}\label{fig:2b}
\vspace{1em}
\end{subfigure}
\begin{subfigure}{0.35\textwidth}
\includegraphics[width=\textwidth]{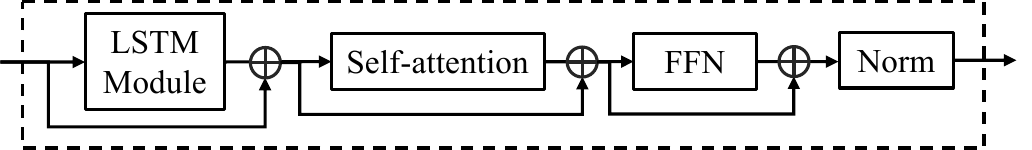}
\caption{LSTM-attention block.}\label{fig:2c}
\vspace{-.5em}
\end{subfigure}
\caption{{Architectures of proposed (a) separator; (b) dual-path block; and (c) LSTM-attention block for intra- and inter-chunk processing}}
\label{fig:2}
\vspace{-.3cm}
\end{figure}

The separator takes the encoder output $\textbf{E}$ as input and produces ${C}$ source representation $\textbf{Z}_{c}\in{\mathbb{R}^{T' \times D_{e}}}$ where ${c\in{\{1, \cdots, C\}}}$. The overall separator architecture is depicted in Fig. \ref{fig:2a}. $\textbf{E}$ is first fed into a linear layer with $D$ output units, and then split into chunks of $K$ frames with a hop size of $\left \lceil K/2 \right \rceil$, following~\cite{Luo2020}.
This operation generates $S$ equal-size chunks $\textbf{U}^s\in{\RR}^{K \times D}$ where ${s\in\{1, \cdots, S\}}$.
All chunks are stacked along a new chunk axis to obtain a three-dimensional tensor $\textbf{U}\in{\RR}^{K \times S \times D}$. 
% The mode-$1$ and mode-$2$ are commonly called intra- and inter-chunk axis, respectively.

\subsubsection{Dual-path processing}

The segmented tensor $\textbf{U}$ is fed into a dual-path processing block depicted in Fig. \ref{fig:2b}:
\begin{align}
\label{eq:2}
\textbf{U${'}$}&=[f_{\text{intra}}(\textbf{U}[:,s,:]), \forall{s}]\in{\RR}^{K \times S \times D},\\
\label{eq:3}
\textbf{U${''}$}&=\text{LN}([f_{\text{inter}}(\textbf{U${'}$}[k,:,:]), \forall{k}]+\textbf{U}),
\end{align}
where ${k\in\{1, \cdots, K\}}$ and $f_{\text{intra}}(\cdot)$, $f_{\text{inter}}(\cdot)$, and $\text{LN}(\cdot)$ are intra-, inter-chunk processing, and layer normalization, respectively.
Instead of utilizing multiple blocks, a single block is employed for dual-path processing in our experiments.
For $f_{\text{intra}}$ and $f_{\text{inter}}$, we improve SepFormer~\cite{Subakan2021} by incorporating an LSTM-augmented architecture, named LSTM-attention block, to better capture local and global contextual information with direct context-awareness~\cite{Chen2020DPTnet}. The LSTM-attention block is illustrated in Fig.~\ref{fig:2c}. The LSTM module consists of a layer normalization~\cite{Ba2016LayerN}, followed by BLSTM~\cite{Hochreiter1997Long} and a linear projection layer to transform back to the feature dimension $D$. Self-attention and feed-forward modules are the same as~\cite{Subakan2021} except for the activation function (we replace the ReLU
%~\cite{Glorot2011Deep}
activation function with GELU).
A residual connection is added after each module, followed by a layer normalization.
Furthermore, in the self-attention module, we incorporate explicit positional information by employing T5-style relative position embeddings with per-head bias~\cite{Raffel2020T5}. Unlike SepFormer~\cite{Subakan2021}, the LSTM-attention block
does not require several transformer layers in each intra- or inter-chunk block.
%Other combinations, including altering the position of the LSTM module or DPTNet-style block~\cite{Chen2020DPTnet} (e.g., BLSTM after multi-head self-attention),
%including 
%resulted in
%%the unstable 
%model divergence during training in our experiments.
We tried to modify the order of the LSTM, self-attention, and FFN modules but did not observe any improvement.

% \begin{figure}
% \centering
% \includegraphics[width=6.56cm]{figs/fig2.jpg}
% \caption{SepLSTMFormer block for intra- and inter-chunk processing.}
% \label{fig:2}
% \end{figure}

\begin{figure}
\centering
\includegraphics[width=0.46\textwidth]{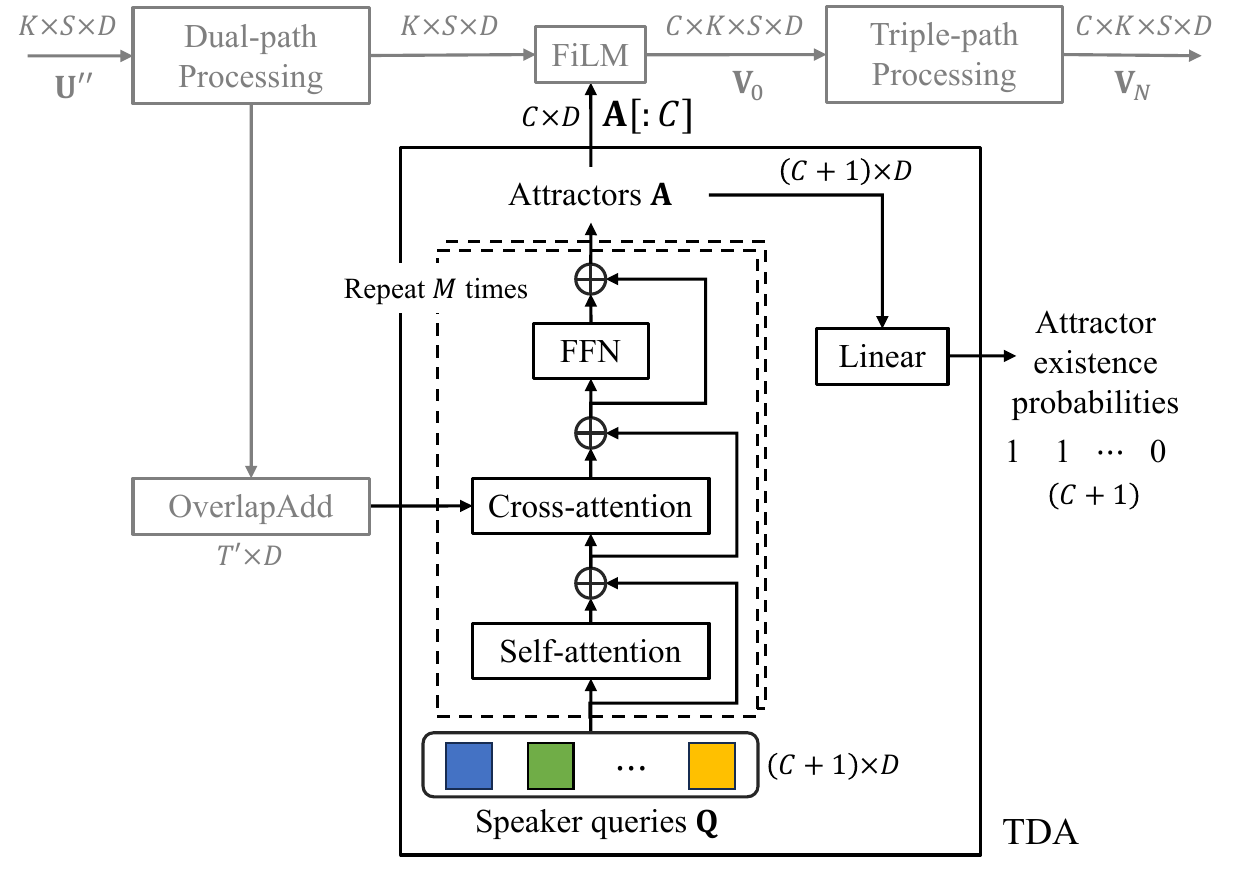}
\vspace{-.5em}
\caption{Transformer decoder-based attractor calculation module.}
\label{fig:3}
\vspace{-.5cm}
\end{figure}

\subsubsection{Transformer decoder-based attractor calculation module}
\label{sssec:tda}
In previous studies that applied traditional EDA to SS~\cite{Chetupalli2022, Maiti2023EENDSS}, an additional layer is needed to reduce the sequence length due to the fact that the LSTM encoder is not well-suited for handling extremely long sequences. In addition, due to the nature of the LSTM encoder-decoder, EDA squashes all the information of input into a single fixed-length vector, which may yield a bottleneck in improving the performance~\cite{Bahdanau2015}. In order to overcome these limitations, inspired by the idea of ~\cite{Carion2020End, Horiguchi2020end}, we propose a TDA calculation module by extending EDA to a transformer-based architecture.

The TDA calculation module, depicted in Fig.~\ref{fig:3}, consists of $M$ transformer decoder layers. During training, the number of speakers $C$ is assumed known, and $C+1$ speaker query embeddings, $\textbf{Q}\in{\RR}^{(C+1) \times D}$, are randomly initialized and learned through the training process. TDA aims to estimate the attractors $\textbf{A}$ based on speaker queries $\textbf{Q}$ and mixture context as follows:
\begin{align}
\label{eq:4}
\textbf{A}&=\text{TDA}(\text{OverlapAdd}(\textbf{U${''}$}),\textbf{Q})\in{\RR}^{(C+1) \times D}.
\end{align}
The context is obtained by employing the overlap-add operator to the dual-path processing output \textbf{U${''}$} in Eq.~\eqref{eq:3} and is utilized as a context in cross-attention calculation, which enables to attend to the entire sequence with length $T'$, transforming each speaker query into a speaker-wise attractor.
Note that the cross-attention operation has low complexity since $C$ is much smaller than $T'$.
% The TDA module consists of $M$ transformer decoder layers and takes a series of learned embeddings known as speaker queries as input. It attends to the full mixture representation, converting each speaker query into a speaker-wise attractor.
% In previous studies that applied traditional EDA to SS tasks~\cite{Chetupalli2022, Maiti2023EENDSS}, an additional layer is needed to reduce the sequence length. This is primarily due to the fact that the LSTM encoder is not well-suited for handling extremely long sequences. Therefore, by replacing LSTM encoder-decoder structure with transformers. TDA can mitigate this bottleneck problem that arises with the use of a fixed-length state vector in EDA by directly attending to entire sequences as the cross-attention operation has low complexity with $C~{\ll}~T'$.
% \cite{Fujita2023NAR} has proposed a method similar to the TDA module in the field of speaker diarization, however, their approach is limited to a fixed number of speakers. In order to handle a flexible number of speakers, 
We adopt masked self-attention in the TDA calculation module to prevent the $c$-th attractor prediction from attending to the next  (${>c}$) speaker queries.
% For instance, in the case of a mixture with two speakers in the mixture, the first two queries are constrained to process those speakers exclusively.
The first $C$ attractors are responsible for speaker identification, while the last ($C+1$)-th attractor is utilized for estimating the speaker non-existence.
% By employing self- and cross-attention mechanisms on these query embeddings, the TDA module comprehensively considers all speaker candidates collectively, utilizing pairwise relationships among them, while retaining the capacity to utilize the entire sequence as context.
The self-attention calculation in the first decoder layer is skipped as in~\cite{Carion2020End}. After the TDA calculation, $C$ speaker-wise attractors (denoted as $\textbf{A}[{:C}]$) are combined with the dual-path processing output $\textbf{U${''}$}$ by feature-wise linear modulation (FiLM) conditioning~\cite{Perez2018film}, to generate 4-D tensor output $\textbf{V}_{0}$:
\begin{align}
\label{eq:5}
\textbf{A}[{:C}]=[\textbf{a}_{1},\cdots,\textbf{a}_{C}]&\in{\RR}^{C \times D},\\
\label{eq:6}
\textbf{V}_{0}=\text{FiLM}(\textbf{U${''}$}, \textbf{A}[:C])&\in{\RR}^{C \times K \times S \times D},
\end{align}
where ${\text{FiLM}}(\textbf{F}, \textbf{d})=\text{Linear}(\textbf{d})\odot\textbf{F}+\text{Linear}'(\textbf{d})$ with two different linear projections of $\textbf{d}$.
%The overall structure of the TDA calculation module is depicted in Fig.~\ref{fig:3}.

\begin{figure}
\centering
\includegraphics[width=0.48\textwidth]{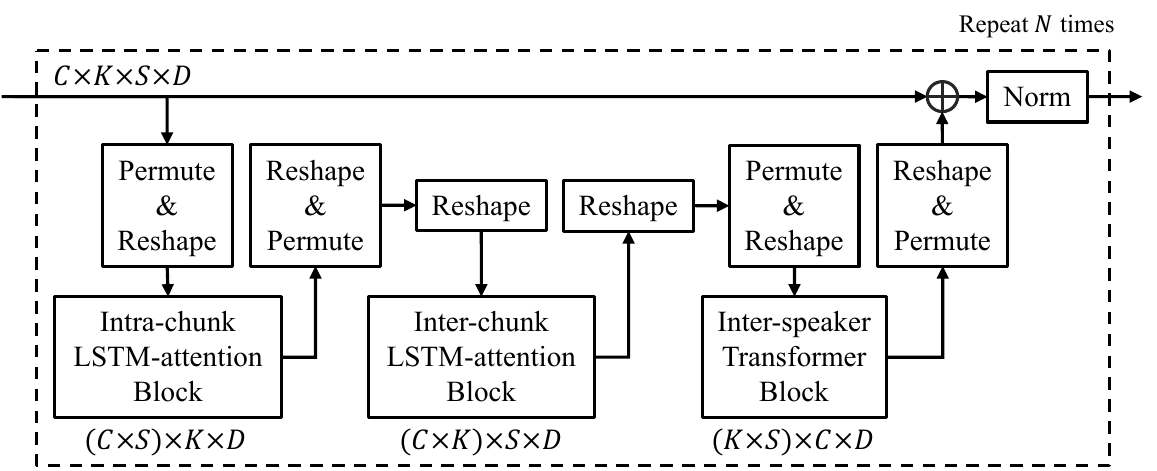}
\vspace{-1.5em}
\caption{Triple-path processing block.}
\label{fig:4}
\vspace{-.5cm}
\end{figure}

\subsubsection{Triple-path processing}
\label{sssec:triple}

Following~\cite{Chetupalli2022}, $\textbf{V}_{0}$ is refined by $N$ triple-path processing blocks which extend the dual-path block with an additional inter-speaker transformer module.
See Fig.~\ref{fig:4} for an illustration.
As in the previous dual-path processing, we use LSTM-attention blocks in the intra- and inter-chunk processing, while for the inter-speaker processing, a simple transformer layer is used since the order of speakers is not irrelevant due to the use of the PIT loss~\cite{Kolbak2017}.
Residual connections are added after each processing block, followed by a layer normalization layer. The triple-path processing block is denoted as follows:
\begin{align}
\label{eq:7}
\textbf{V}_{n-1}^{'}=&[f_{n, \text{intra}}(\textbf{V}_{n-1}[c,:,s,:]), \forall{c, s}],\\
\label{eq:8}
\textbf{V}_{n-1}^{''}=&[f_{n, \text{inter}}(\textbf{V}_{n-1}^{'}[c,k,:,:])\forall{c, k}],\\
\label{eq:9}
\textbf{V}_{n}=\text{LN}(&[f_{n, \text{speaker}}(\textbf{V}_{n-1}^{''}[c,:,:,:]),\forall{c}]+\textbf{V}_{n-1}),
\end{align}
where ${c\in\{1, \cdots, C\}}$ and $f_{n,\text{intra}}(\cdot)$, $f_{n,\text{inter}}(\cdot)$, and $f_{n,\text{speaker}}(\cdot)$ are respectively intra-chunk, inter-chunk, and inter-speaker processing in block $n\in\{1,\cdots, N\}$.
\par After the final triple-path processed output $\textbf{V}_{N}$ is obtained, the overlap-add operation is employed on the $S$ chunks, forming an output of size ${\mathbb{R}^{C \times T' \times D}}$, which has the same sequence length as the encoder output, $\textbf{E}$. Then, layer normalization followed by a feed-forward layer with $D_{e}$ output units is applied, producing $\textbf{Z}_{N,c}\in{\mathbb{R}^{T' \times D_{e}}}$.

% \vspace{-.2cm}
\subsection{Decoder}
% \vspace{-.2cm}
\label{ssec:decoder}
% 디코더에서는 분리한걸 음원으로 복원하는 작업을 하며, 이는 트리플 패쓰 프로세싱을 통해 진행된다.
The decoder reconstructs each source waveform $\hat{\textbf{y}}_{c}$ from $\textbf{Z}_{N,c}$ using the transposed convolution version of the encoder (i.e., with a kernel size of $L$ samples and a stride size of $L/2$ samples) as follows:
\begin{align}
\label{eq:10}
\hat{\textbf{y}}_{N,c}=\text{TransposedConv1D}(\textbf{Z}_{N,c}).
\end{align}

% \vspace{-.3cm}
\subsection{Loss Functions}
\label{ssec:loss}
\par Inspired by~\cite{Zeghidour2020, Nachmani2020}, a multi-scale loss technique is applied to our training objective. The reconstruction loss is computed from each output of triple-path processing block $\textbf{V}_{n}$ in Eq.~\eqref{eq:9}.
Then, the average over all the triple-path blocks is used as the loss.
The reconstruction loss is defined by SI-SDR~\cite{Luo2018TasNet, LeRoux2018a} as follows:
\begin{align}
\label{eq:11}
\mathcal{L}_{\text{recon}}=-\frac{1}{N}\sum_{n=1}^{N}\max_{\pi \in{{\Pi}_{C}}}{\frac{1}{C}\sum_{c=1}^{C}\textrm{SI-SDR}(\textbf{y}_{c}, \hat{\textbf{y}}_{n,\pi(c)})},
\end{align}
where $N$ denotes the number triple-path blocks, $\Pi_{C}$ is the set of all the permutations over $C$ speakers and
utterance-level PIT~\cite{Kolbak2017} is used to address the label-permutation problem,
%the best permutation is calculated and chosen only if $n=N$ to update all the trainable parameters through backpropagation, which is denoted as the utterance-level PIT~\cite{Kolbak2017}. 
and $\hat{\textbf{y}}_{n,c}$ is the estimated $c$-th source obtained from the $n$-th triple-path processing block output. Note that only the final output $\hat{\textbf{y}}_{N,c}$ is used at inference time.
\par The training objective of the attractor existence probabilities is based on binary cross-entropy:
\begin{align}
\label{eq:12}
\mathcal{L}_{\text{attractor}}=\text{BCE}\big(\textbf{a}, \sigma(\text{Linear(}\textbf{A}))\big),
\end{align}
where $\textbf{A}$ is a set of attractor vectors in Eq.~\eqref{eq:4}, $\textbf{a}=[1, \cdots, 1, 0]^{\T}\in{\mathbb{R}^{C+1}}$, and $\sigma(\cdot)$ is the sigmoid activation function. $\text{BCE}(\cdot)$ denotes the binary cross entropy loss between target and input logits. Our final loss is $\mathcal{L}_{\text{total}}=\mathcal{L}_{\text{recon}} + \mathcal{L}_{\text{attractor}}$.

\section{Experimental Settings}
\label{sec:exsetting}
% We describe the experimental setup of the SepTDA for performance evaluation.
% We describe benchmark datasets, the experimental conditions, and the evaluation metrics for a fair comparison.
% \vspace{-.2cm}
\subsection{Datasets}
\label{ssec:datasets}
We evaluate SepTDA on popular SS benchmark datasets, WSJ0-\{2,3,4,5\}mix~\cite{Hershey2016, Isik2016, Nachmani2020}. WSJ0-\{2,3\}mix~\cite{Hershey2016, Isik2016} are the most widely used benchmark datasets in monaural speaker-independent SS tasks, which respectively consist of 2- and 3-speaker mixtures simulated based on clean speech signals in the WSJ0 corpus.
%Later, the dataset of speech mixtures of \{4, 5\} utterances is released by \cite{Nachmani2020}.
Later, WSJ0-\{4,5\}mix with mixtures of 4 and 5 utterances are released by \cite{Nachmani2020}.
Unlike WSJ0-\{2,3\}mix, WSJ0-\{4,5\}mix have some mixtures with utterances from the same speaker.
In these datasets, the relative level of a speech signal is randomly chosen from the range $[0, 5]$\,dB.
Each dataset has $20000$, $5000$, and $3000$ mixtures for training, validation, and testing, respectively, and the training and test sets do not share any speakers.
The sampling rate is $8$~kHz.

% wsj0-2mix는 싱글채널 스피커 인디펜던트 ss 분야에서 가장 널리 쓰이는 벤치마크 데이터 셋이며, 이는 wsj데이터셋의 클린 스피치를 이용해서 시뮬레이트된 믹스쳐이다.
% 이 데이터셋이 확장되어 3,4,5믹스가 공개되어있다. 
% 스피커 신호는 상대적인 크기로 랜덤하게 0-5 db 사이로 섞여있다.
% 모든 데이터셋은 2만개의 학습셋, 5천개의 밸리데이션, 3천개의 테스트 셋으로 구성되어있고, 학습, 테스트 셋은 겹치지 않는 스피커로 구성되어있다.
% 우리 실험은 이 데이터셋의 8k 샘플링 레이트를 기준으로 수행되었다.
% 믹스쳐는 발화 중간의 사일런스를 제외하고는 전체가 오버랩 되어있는 min 세팅으로 진행하였다.

% is sampled uniformly from the range $[-5, 5]$\,dB.
% The sampling rate is 8 kHz. %with an energy level randomly sampled from the range $[-5, 5]$ dB.
% \vspace{-.2cm}
\subsection{Experiment configurations}
\label{ssec:exconf}
The encoder/decoder kernel size $L$ in Section \ref{sec:encoder} and \ref{ssec:decoder} is set to 16 with a stride size of 8 samples.
The feature dimensions $D_{e}$ and $D$ in Section \ref{ssec:decoder} are set to 256 and 128, respectively. For dual-path chunking, our model processes chunks of length $K=96$ in Section \ref{sec:separator} with 50\% overlap. The number of TDA decoder layers $M$ in Section \ref{sssec:tda} and triple-path blocks $N$ in Section \ref{sssec:triple} is set to 2 and 8 respectively, and the number of hidden units in BLSTM is set to be $256$ in each direction. We use 4 attention heads and an expansion factor of 4 for the feed-forward module.
\par Following~\cite{Chetupalli2022}, SepTDA is trained with three different dataset combinations: 1) WSJ0-2mix only, 2) WSJ0-\{2,3\}mix combined, and 3) WSJ0-\{2,3,4,5\}mix combined. We denote the three models as $\text{SepTDA}_2$, $\text{SepTDA}_{2/3}$, and $\text{SepTDA}_{[2-5]}$, respectively. In contrast to \cite{Chetupalli2022}, all the models are trained from scratch without using any data augmentation.% strategy, e.g., speed perturbation or dynamic mixing.

The AdamW optimizer~\cite{Loshchilov2018Decoupled} is used with an initial learning rate of $4\times 10^{-4}$. The learning rate is halved when a validation loss has stopped improving for 5 successive epochs. We clip the $L_{2}$ norm of the gradients to 5. The batch size is set to 2 and the length of each training example is set to 4 seconds during training.
% Automatic mixed precision is applied to speed up training.

% \vspace{-.2cm}
\subsection{Evaluation metrics}
%To evaluate the separation performance of the proposed method
SI-SDR improvement ($\Delta$SI-SDR)~\cite{LeRoux2018a} and SDR improvement ($\Delta$SDR)~\cite{Vincent2006a} are used as the evaluation metrics. Following \cite{Chetupalli2022}, we also consider two scenarios when evaluating SepTDA$_{2/3}$ and SepTDA$_{[2-5]}$, depending on whether the number of speakers is known or not. When it is unknown, the estimated number of speakers, $\hat{C}$, may not necessarily match the ground truth $C$. When there is an overestimation of the number of speakers (i.e., $\hat{C}>C$), we select the estimated sources corresponding to the initial $C$ speaker queries. In cases of an underestimation in the number of speakers (i.e., $\hat{C}<C$), a silence signal (i.e., an all-zeros signal) is used as the estimate for the speakers that were not estimated as in \cite{Chetupalli2022}.
Since the SI-SDR and SDR metrics are undefined if the reference signal is all-zero (i.e., $\text{log}(0)$ is undefined), an epsilon value of $10^{-8}$ is used for the final log calculation, resulting in an SI-SDR value of $-80$ dB.
% unlike \cite{Chetupalli2022} which assigns an all-zero signal to each of the $C-\hat{C}$ sources, a Gaussian random noise signal is used as the estimate for the speakers that were not estimated, since the SI-SDR and SDR metrics are undefined if the reference signal is all-zero.In cases of an underestimate in the number of speakers ($\hat{C}<C$), a silence signal (i.e., an all-zeros signal) is used as the estimate for the speakers that were not estimated. (e.g., $\text{log}(0)$).
The average mixture SI-SDR scores of 2 to 5mix are respectively $0$, $-3.2$, $-5.0$, and $-6.3$ dB.

\section{Evaluation Results}\label{results}
% 이번 섹션에서는 우리의 실험결과를 보여주고, 다른 결과와 어떻게 다른지 분석해보도록 한다.
We analyze SepTDA by training it on different combinations of datasets and comparing its performance with previous studies.
Then, we discuss the advantages of SepTDA through ablation studies.

\subsection{Results on WSJ0-2mix}
% 먼저, 두 화자에 대한 믹스쳐에 대한 실험 결과를 비교해보았다.
% 표2에 기존 알고리즘과 우리가 wsj0-2mix에서 학습한 결과를 수록하였다.
% 공평한 비교를 위해서 dynamic mixing[]이나 data augmentation을 제외하고, 동일한 데이터셋을 사용한 결과만을 수록하였다.
% 그 결과 우리가 제안한 결과는 현재 나온 분리 모델중에 가장 좋은 성능을 보이는걸 확인할 수 있었다. 
% 우리 모델은 sepfomer에 lstm이 추가 되고, eda모듈을 transformer를 이용해서 모든 시퀀스를 다 확인하는 방법을 사용했다는 점에서 기존 연구와 다르다고 할 수 있다. 

%We analyze the experimental results on the two-speaker mixture dataset, WSJ0-2mix. 
Table \ref{table:1} compares the separation performance of our models on WSJ0-2mix with previous systems in terms of $\Delta$SI-SDR and $\Delta$SDR.
% The results of the previous algorithms and the proposed model are shown in Table \ref{table:2}.
% For a fair comparison, we only include results using the same dataset. %, excluding dynamic mixing or any data augmentation~\cite{Subakan2021}.
% Results demonstrate that our model performs the best among the existing separation models.
The SepTDA$_2$ model outperforms the current best reported in the TF-GridNet paper \cite{Wang2023TFGridNet2} by achieving a $\Delta$SI-SDR of $23.7$ dB. 
% We have also achieved further performance improvement by reducing the kernel size $L$ from 16 to 12, resulting in 24.0 dB, at the expense of computational cost (see the last row in Table \ref{table:3}).
The performance is further improved to $24.0$ dB by reducing the kernel size $L$ introduced in Section \ref{sec:encoder} and \ref{ssec:decoder} from $16$ to $12$ samples, at the cost of a marginal increase in the amount of computation (see the last row in Table \ref{table:3}).
% Our model differs from earlier studies in that we used a transformer in the EDA module to see all sequences and added LSTM to SepFormer. 

\begin{table}[t]
\scriptsize
\centering
\sisetup{table-format=2.2,round-mode=places,round-precision=2,table-number-alignment = center, detect-weight=true, detect-inline-weight=math}
\caption{Comparison with previous models on WSJ0-2mix. TF and Q-Dual denote time-frequency and quasi-dual-path, respectively. ‘*’ denotes the utilization of data augmentation (e.g., dynamic mixing or speed perturbation).}
\vspace{-0.2cm}
\label{table:1}
\setlength{\tabcolsep}{1pt}
\begin{tabular}{
l
cc
S[table-format=3.1,round-precision=1]
S[table-format=2.1,round-precision=1]
S[table-format=2.1,round-precision=1]
}
\toprule

Models & Domain & Path & {\#params (M)} & {$\Delta$SI-SDR (dB)}  & {$\Delta$SDR (dB)}\\

\midrule

% Conv-TasNet~\cite{Luo2019conv} & Time & Single & 5.1 & 15.3 & 15.6 \\
DPRNN~\cite{Luo2020} & Time & Dual & 2.6 & 18.8 & 19.0 \\
Gated DPRNN~\cite{Nachmani2020} & Time & Dual & 7.5 & 20.1 & 20.4 \\
DPTNet~\cite{Chen2020DPTnet} & Time & Dual & 2.7 & 20.2 & 20.6 \\
SepFormer~\cite{Subakan2021} & Time & Dual & 26.0 & 20.4 & 20.5 \\
% Sandglasset~\cite{Lam2021} & Time & Dual & 2.3 & 20.8 & 21.0 \\
Wavesplit~\cite{Zeghidour2020} & Time & Single & 29.0 & 21.0 & 21.2 \\
% TFPSNet~\cite{Yang2022TFPSNet} & TF & Dual & 2.7 & 21.1 & 21.3 \\
% SFSRNet~\cite{Rixen2022} & Time & Dual & 59.0 & 22.0 & 22.1 \\
QDPN~\cite{Rixen2022QDPN} & Time & Q-Dual & 200.0 & 22.1 & {-} \\
SepEDA${_2}$$^*$~\cite{Chetupalli2022} & Time & Triple & 12.5 & 21.2 & 21.4 \\
% SepEDA${_{2/3}}$~\cite{Chetupalli2022} & Time & Triple & 12.5 & 21.5 & 21.7 \\
MossFormer(L)$^*$~\cite{Zhao2023Moss} & Time & Single & 42.1 & 22.8 & {-} \\
TF-GridNet~\cite{Wang2023TFGridNet2} & TF & Dual & 14.5 & 23.5 & 23.6 \\
\midrule
SepTDA${_2}$ & Time & Triple & 21.2 & \bfseries 23.7108 & \bfseries 23.4624 \\
~~~with $L=12$ & Time & Triple & 21.2 & \bfseries 24.0458 & \bfseries 23.9428 \\
\bottomrule
\end{tabular}
\vspace{-0.4cm}
\end{table}

% 다화자에 대한 분리시, 화자수 추정 정확도에 대해서 알아보았다.
% 화자 추정 정확도는 표3과 같이 나왔는데, 이는 기존연구 [22, chet2023] 보다도 높으며, 추후 diarization, 실시간 화자 분리와의 매끄러운 연계의 가능성을 보여준다.

% \begin{table}[t]
% \scriptsize
% \centering
% \caption{Performance of speaker number estimation accuracy}
% \label{tbl:nspkraccuracy}
% \begin{tabular}{ccccc}
% \toprule
% \multirow{2}{*}{Number of sources} & \multicolumn{4}{c}{Estimated number of sources} \\
%                                    & 2              & 3              & 4       & 5      \\
%                                    \midrule
% 2                                  &    99\%            &      -          &     -     &  -   \\
% 3                                  &       1\%         &    98\%            &    -      &  -   \\
% 4                                  &        -        &       2\%         &   97\%      &   -   \\
% 5                                  &       -         &      -          &    3\%     &  96\%    \\
% \bottomrule
% \end{tabular}
% \end{table}

% \vspace{-0.25cm}
\subsection{Results on WSJ0-$C$mix}

% 다음 우리는 flexible 숫자의 화자 믹스쳐에 대한 실험 결과를 얘기한다.
% 비교를 위해서 우리는 두가지 모델을 추가로 학습하였다. 
% 한가지는 2/3에 대해서 학습한 모델, 다른 한가지는 2-5 화자를 모두 학습한 모델이다.
% 실험 결과는 표4에 표현하였다. 
% 표를확인해보면 우리 모델의 성능이 가장 좋은 것을 알 수 있다. 
% 화자의 숫자가 정해진 경우가 조금 더 조건이 완화된 경우지만, 우리 실험은 가장 최근에 나온 연구보다 성능이 좋은것을 확인할 수 있다.
% 제안한 내용이 다화자에도 잘 적용이 되는것을 알수 있다.

% 다음은 2명 이상의 믹스쳐에 대해서 음성을 분리한 결과를 분석해보았다.
% 기존 모델과 비교를하기 위해서 우리는 두 방식으로 학습을 진행하고 그 결과를 표에 표시하였다.
% 표 윗부분은 2,3명까지 분리가 가능한 모델들과, 2,3으로 학습한 proposed 2/3이며 표 아랫부분은 2-5화자까지 분리할 수 있는 모델들과 2-5까지 학습한 proposed [2-5]를 비교하였다.
% 음성 분리 모델들은 그특징에 따라서 분리할 화자 숫자를 미리 알고 있는 경우와 섞인 화자의 숫자를 알려주지 않은 경우로 나눌수 있으며, 화자 숫자를 알고 있는 경우는 표에 * 표시를 하였다.

% Next, we analyze the results of SS for mixtures with more than two speakers.
% To compare with the existing models, we trained the proposed models with two methods: $\text{SepTDA}_{2/3}$ and $\text{SepTDA}_{[2-5]}$.
% The results are shown in Table \ref{table:2} where the upper part of the table compares models that can separate up to 3 speakers and the lower part of the table compares models that can separate up to 2 to 5 speakers, they are compared with $\text{SepTDA}_{2/3}$ and $\text{SepTDA}_{[2-5]}$, respectively.
% The SS models can be classified into two categories according to the characteristics: those where the number of speakers to be separated is known in advance($\star$) and those where it is unknown.
Table~\ref{table:2} compares the results of SepTDA with previous systems which can separate mixtures with up to 3 or 5 speakers. 
Models marked as flexible $C$ (by using \cmark) are those that can specify the number of speakers to be separated, while the others can only separate a mixture into a fixed number of sources that they were trained with and require multiple models.
For the models where the number of speakers is unknown, marked with `$^\star$', the speaker counting accuracy results are also reported.
When the estimated attractor existence probability exceeds $0.5$, we count it as an indication of the presence of a speaker, which is similar to Algorithm 1 of \cite{Chetupalli2022}.
% The cases with the known number of speakers are indicated with $\star$.

% 2,3명이 분리 가능한 모델들을 봤을때, 우리 모델이 가장 높은 성능을 보이는 것을 확인 할수 있었다.
% 화자 숫자를 모르는 경우, 알고 있는 경우보다 성능이 조금 떨어지지만 그 경우에도 동일 데이터셋으로 학습한 기존 연구보다도 가장 좋은 성능을 보여주었다.
% 5명까지 분리가능한 연구들과 비교한 경우에도 가장 높은 성능을 보였다.
% 특히, 분리할 화자수가 정해진 경우에는 분리할 화자 수가 늘어남에 따라 줄어드는 성능폭이 기존 연구들 대비 굉장히 적으며, 5명을 분리하는 경우에도 기존의 2명 분리하는 경우와 큰 차이가 나지 않았다.

For the cases of 2 and 3 speakers, we observe that SepTDA outperforms the previous best systems by a large margin.
Although the performance with an unknown number of speakers is slightly worse than that with a known number of speakers, it is better than the previous studies trained on the same dataset. SepTDA also performs the best when compared to studies that can separate up to 5 speaker sources. In particular, when the number of speakers is known, the performance degradation as the number of speakers increases is minimal compared to previous studies, and even for the 5mix dataset, $\Delta$SI-SDR is not significantly different from the 2mix results of the previous studies. 
For an unknown number of speakers, although the speaker counting accuracy is lower than the previous SepEDA, we achieve much higher $\Delta$SI-SDR.
We could potentially improve speaker counting accuracy by stacking more layers instead of a single linear projection layer in Eq.~\eqref{eq:12}, or by employing fine-tuning. However, we will leave these possibilities for future work.

\begin{table}[t]
\scriptsize
\centering
\sisetup{table-format=2.2,round-mode=places,round-precision=2,table-number-alignment = center,detect-weight=true,detect-inline-weight=math}
\caption{{$\Delta$SI-SDR (dB)} comparison on WSJ0-\{2,3,4,5\}mix. ‘$^\star$’ denotes situations with an unknown number of speakers. Numbers in parentheses represent speaker counting accuracy in percentage.}
\vspace{-0.2cm}
\label{table:2}
\setlength{\tabcolsep}{2.7pt}
\begin{tabular}{lcr@{\hspace{-0ex}}lr@{\hspace{-0ex}}lr@{\hspace{-0ex}}lr@{\hspace{-0ex}}l}
\toprule
Models   & Flexible $C$          & 2    & & 3    & & 4  &  & 5 &   \\ \toprule
SepFormer~\cite{Subakan2021}       & \xmark & 20.4 & & 17.6 & & -  &  & - &   \\
Wavesplit~\cite{Zeghidour2020}     & \xmark & 21.0 & & 17.3 & & -  &  & - &   \\
MossFormer(L)~\cite{Zhao2023Moss}  & \xmark & 22.8 & & 21.2 & & -  &  & - &   \\
SepEDA$_{2/3}$~\cite{Chetupalli2022} & \cmark & 21.5 & & 19.9 & & -  &  & - &   \\
SepTDA$_{2/3}$   & \cmark & \bfseries 24.0 & & \bfseries 23.7 & & - &   & - &   \\
SepTDA$^{\star}_{2/3}$   & \cmark & \bfseries 24.0 & $_\text{(99.97)}$ & \bfseries 22.8 &$_\text{(97.93)}$ & - &   & -  &  \\ \midrule 
Recursive SS~\cite{Takahashi2019} & \cmark & 14.8 & & 12.6 & & 10.2& & -  &  \\
Gated DPRNN~\cite{Nachmani2020} & \xmark & 20.1 & & 16.9 & & 12.9 & &10.6 &\\
Gated DPRNN$^{\star}$~\cite{Nachmani2020} & \xmark & 18.6 & & 14.6 & & 11.5& & 10.4 &  \\
SepEDA$_{[2-5]}$~\cite{Chetupalli2022} & \cmark & 21.1 & & 18.6 & & 14.7 & &12.1 &\\
SepEDA$^{\star}_{[2-5]}$~\cite{Chetupalli2022}       & \cmark & 21.1 & $_\text{(99.80)}$ & 18.4 & $_\text{(97.00)}$ & 14.4&$_\text{(90.17)}$          & 11.6&$_\text{(96.87)}$          \\
SepTDA$_{[2-5]}$ & \cmark & \bfseries 23.6 & & \bfseries 23.5 & & \bfseries 22.0 & &\bfseries 21.0 & \\
SepTDA$^{\star}_{[2-5]}$ & \cmark & \bfseries 23.6 & $_\text{(99.90)}$ & \bfseries 22.1 & $_\text{(95.93)}$ & \bfseries 19.5 &$_\text{(90.10)}$ & \bfseries 16.9 &$_\text{(83.23)}$ \\ \bottomrule
\end{tabular}
\vspace{-.15cm}
\end{table}

% silence for under-estimated
% model trained on 2&3mix (unknown / known)
% 24.03847557359934 / 24.04755046548446
% 22.829202948729197 / 23.69188508494695
% model trained on 2-5mix (unknown / known)
% 23.56372574234009 / 23.59868256664276 / 99.90
% 22.131903812964758 / 23.52947906422615 / 95.93
% 19.537836741288505 / 21.959747952461242 / 90.10
% 16.91246817692121 / 20.939143766641617 / 83.23

% model trained on 2&3mix (unknown / known)
% 24.03847556549311 / 24.04755047518015
% 23.05981208149592 / 23.69188541142146
% model trained on 2-5mix (unknown / known)
% 23.56258969183763 / 23.594017028212548 / 99.90
% 22.92241156943639 / 23.540998038212457 / 95.90
% 20.950214699824652 / 21.974059567689896 / 89.83
% 19.601707660039267 / 20.95865792878469 / 82.97

% \subsection{Performance of speaker counting}

\begin{table}[t]
\scriptsize
\centering
\sisetup{table-format=2.2,round-mode=places,round-precision=2,table-number-alignment = center,detect-weight=true,detect-inline-weight=math}
\caption{Ablation results on LSTM-attention block based on WSJ0-2mix.}
% \vspace{-0.2cm}
\label{table:3}
\setlength{\tabcolsep}{1.5pt}
\begin{tabular}{
r
l
c
S[table-format=2.1,round-precision=1]
S[table-format=2.1,round-precision=1]
S[table-format=2.1,round-precision=1]
}
\toprule

\#&Models & Masking/Mapping & {\#params (M)} & {$\Delta$SI-SDR (dB)} & GMAC/s \\

\midrule
1&DPRNN~\cite{Luo2020} & Masking & 2.6 & 18.8 &  42.2  \\ 
2&Gated DPRNN~\cite{Nachmani2020} & Mapping & 7.5 & 20.1 &  31.54  \\
3&SepFormer~\cite{Subakan2021} & Masking & 26.0 & 20.4 &  59.5 \\
4&TF-GridNet~\cite{Wang2023TFGridNet2} & Mapping & 14.5 & 23.5 & 231.1 \\

\midrule
5&LSTM-attention & Masking & 17.0 & 21.36 &  36.1975 \\
6&LSTM-attention & Mapping & 17.0 & \bfseries 22.04 & 36.1975 \\
7&~~~w/o self-attention  & Mapping & 16.0 & 20.88 & 34.05 \\
8&~~~w/o LSTM  & Mapping & 13.0 & 19.98 & 31.1075 \\
\midrule
9&SepTDA$_2$ & Mapping & 21.2 & \bfseries 23.7 & 81.0475 \\
10&~~~with $L=12$ & Mapping & 21.2 & \bfseries 24.0458 & 107.7475 \\
\bottomrule
\end{tabular}
% \vspace{-0.4cm}
\end{table}

\subsection{Ablation results on LSTM-attention block}

% 우리 모델의 효과가 어디서 오는지 확인하기위해서 어블레이션 스터디를 진행했다. 그 결과는 표3에 있다.
% 베이스라인으로 비교하는 모델은 ~~~ 이고, ~~를 같이 표시하였다.
% 제안하는 모델은 동일한 조건에서 비교하기 위해서 wsj0-2mix로 학습을 하여고,lstmsepforemr에서 한가지씩 제거하면서 성능을 확인하였다.

To study the effects of each module in SepTDA, we conduct an ablation study with a comparison with various baseline models.
The results are reported in Table~\ref{table:3}. 
DPRNN~\cite{Luo2020}, Gated DPRNN~\cite{Nachmani2020}, and SepFormer~\cite{Subakan2021} are chosen as the baseline models because their architecture highly relies on BLSTM or Transformer layers. In addition, TF-GridNet~\cite{Wang2023TFGridNet2}, the current best model, is included and the computation cost is reported in terms of GMAC/s as in~\cite{Wang2023TFGridNet2}.

% To analyze the proposed methods, we study the effects of each module by removing it within the LSTMSepFormer block based on the WSJ0-2mix dataset.
% To study the effect of 
For a fair comparison with the baseline models, the proposed models are trained using WSJ0-2mix.
For LSTM-attention (from row 5 to 8), the TDA module and triple-path processing are excluded for a fair comparison, and $8$ dual-path processing blocks are stacked as in other dual-path approaches~\cite{Luo2020, Nachmani2020, Subakan2021}.
For row 5, masking is performed in the encoded embedding space and ReLU is used for encoder activation and mask estimation as in~\cite{Subakan2021}.

From the results in row 5 and 6, direct mapping shows better results than the masking method.
This aligns with the findings in TF-GridNet~\cite{Wang2023TFGridNet, Wang2023TFGridNet2}.
The model in row 6 yields the most favorable separation performance while achieving a harmonious balance of model size between Gated DPRNN and SepFormer. Row 7 and 8 also show the advantages of using the self-attention module and the LSTM module, respectively.
Note that, in the model of row 8, the embedding dimension $D$ introduced in Section \ref{sec:separator} is increased to $256$ in order to maintain the total number of parameters as similar as possible, and the resulting model can be seen as a lightweight version of SepFormer.

% \vspace{-.25cm}
\section{Conclusions}\label{conclusion}
% 이 논문에서 우리는 클린 스피커 믹스쳐에 대해서 sota 성늘을 보이는 모델을 선보였다.
% 2명 믹스쳐에서 제일 성능이 좋았던 방법[] 보다 성능이 좋았으며, 화자 수를 미리 알려주지 않은 경우에 대해서도 더 좋은 성응을 보이는 걸 확인했다.
% 우리가 제안하는 모델은 time domain에서  transformer 구조를 사용하지만 lstm을 앞쪽에 배치하여 --- 효과를 유도하였다.
% 또한 여러 화자에 대한 추정을 하기위해서 transformer기반의 개선된 eda를  사용해서 기존 방법 대비 더 많은 정보를 볼 수 있도록하였고,
% 이를 triple -path를 통해 활용하는 모델을 제안하였다.
% 최근 몇년간의  두 화자 분리  ss에서는 귀로는 큰차이를 느낄수 없을 정도로 성능이 고도화 됐었다.
% 우리는 더 많은 화자의 분리에서도 이질감을 느끼기 어려울 정도로 좋은 성능이 나올수 있다는 점을 확인하였다.
We have proposed SepTDA, a monaural speech separation model that can handle mixtures with an unknown number of speakers. We have proposed an improved LSTM-attention block to the previous dual- and triple-path processing blocks.
In addition, the proposed TDA module infers the relations of a set of learned speaker queries and the mixture context and directly generates individual attractors for each speaker.
By combining these strengths together, SepTDA achieves state-of-the-art results on all the datasets examined in this paper.
%, WSJ0-\{2,3,4,5\}mix. 
Even when the number of speakers is unknown, SepTDA shows strong performance and generalizability at separating mixtures with up to 5 speakers.

\vfill\pagebreak

\bibliographystyle{IEEEtran}
% reference 
{\footnotesize
\bibliography{references.bib}
}

% 2,38 둘다 (다른) 딥클러스터링 - 데이터셋 때문에 하나 마저 추가한거같은데 딴데도 둘다쓰는게 나을까요

\end{document}